\documentclass[10pt,twocolumn,twoside]{IEEEtran} 
\usepackage{amsfonts}
\usepackage[cmex10]{amsmath}
\usepackage{amssymb}
\usepackage{bm}
\usepackage{comment}
\usepackage{graphicx}
\usepackage{color}
\newtheorem{problem}{Problem}
\newtheorem{definition}{Definition}
\newtheorem{theorem}{Theorem}
\newtheorem{lemma}{Lemma}
\newcommand{\argmin}{\mathop{\rm arg~min}\limits}
\newcommand{\esssup}{\mathop{\rm ess~sup}\limits}

\def\Lmath{{\mathcal L}}
\def\R{{\mathbb R}}
\DeclareMathOperator{\prox}{prox}

\usepackage{cite}

\ifCLASSINFOpdf
\else
\fi
\begin{document}
%
\title{Discrete-Valued Control by Sum-of-Absolute-Values Optimization}
%
%
%

\author{Takuya~Ikeda,~
        Masaaki~Nagahara,~\IEEEmembership{Senior~Member,~IEEE},
        ~Shunsuke~Ono,~\IEEEmembership{Member,~IEEE}%
\thanks{T. Ikeda and M. Nagahara are with Graduate School of Informatics, Kyoto University, Japan.
S. Ono is with Imaging Science and Engineering Laboratory, Tokyo Institute of Technology.
e-mail: 
{\tt{ikeda.t@acs.i.kyoto-u.ac.jp}} (T.~Ikeda), 
{\tt{nagahara@ieee.org}} (M.~Nagahara),
{\tt{ono@isl.titech.ac.jp}} (S.~Ono).}
\thanks{Manuscript received April 19, 2005; revised September 17, 2014.}}

%
%

\markboth{Journal of \LaTeX\ Class Files,~Vol.~13, No.~9, September~2014}%
{Shell \MakeLowercase{\textit{et al.}}: Bare Demo of IEEEtran.cls for Journals}
%



\maketitle

\begin{abstract}
In this paper, we propose a new design method of discrete-valued control
for continuous-time linear time-invariant systems based on sum-of-absolute-values 
(SOAV) optimization.
We first formulate the discrete-valued control design
as a finite-horizon SOAV optimal control, which is an extended version of $L^1$ optimal control.
We then give simple conditions that guarantee the existence, discreteness, and uniqueness
of the SOAV optimal control.
Also, we give the continuity property of the value function, by which
we prove the stability of infinite-horizon model predictive SOAV control systems.
We provide a fast algorithm for the SOAV optimization
based on the alternating direction method of multipliers
(ADMM), which has an important advantage in real-time control computation.
A simulation result shows the effectiveness of the proposed method.
\end{abstract}


%
\IEEEpeerreviewmaketitle

\section{Introduction}
Discrete-valued control is a control mechanism that
achieves control objectives (e.g. stability) with control inputs
taking values in a finite alphabet
(e.g. bang-bang control: 1-bit control taking $\pm 1$).
Discrete-valued control has a significant advantage in networked control
in which control signals are quantized and transmitted through networks
(see e.g. \cite{BemHeeJoh});
since discrete-valued control signals need not be quantized,
no quantization error may occur.
Also, discrete-valued control has important applications in
DC-DC conversion \cite{PatProZirMak03},
class D amplifier \cite{GeCha09},
hybrid power system \cite{WooRehLox12},
train control \cite{How00},
hormone therapy \cite{TanHirGolBruAih10},
to name a few.

A standard design method for discrete-valued control is
mixed-integer programming \cite{BemMor99}.
Although this directly gives discrete-valued control,
this method requires heavy computation,
and hence it can be used only for relatively slow plant
such as a gas supply system reported in \cite{BemMor99}.
A more tractable method is dynamic quantization proposed in
\cite{AzuSug08a,AzuSug08b}.
In this approach, a dynamic quantizer is designed such that
the quantizer mimics the ideal (i.e. no quantization) continuous output,
and the state space representation of the dynamic quantizer is given 
in a closed form.
This method, however, assumes an infinite alphabet (e.g. the set of integers, $\mathbb{Z}$).
Another approach is the control parametrization enhancing transform
proposed by \cite{LeeTeoRehJen99},
in which the optimal switching times of a piecewise-constant (i.e. discrete-valued)
control input are computed.
This approach assumes that the number of switching is previously known,
which is in practice hard to obtain.

Alternatively, we propose a novel method for discrete-valued control
based on the idea of the {\em sum-of-absolute-values} (SOAV) optimization
\cite{Nag15}.
The proposed optimal control, which we call the
{\em SOAV optimal control},
is an extended version of $L^1$ optimal control
\cite{NagQueNes16} (also known as the minimum fuel control \cite{AthFal}).
The SOAV optimization is convex and hence the solution can be obtained efficiently.
In fact, as shown in Section \ref{subsec:ADMM}, the optimization is
solved, after time-discretization, 
by the {\em alternating direction method of multipliers} (ADMM)
\cite{ADMM1,DRS2,ADMMBoyd}, which is a simple but much faster 
algorithm for large scale problems than the standard interior point method
\cite[Chap 11]{BoyVan}.

For theoretical analysis,
we prove the existence, discreteness,
and uniqueness of the (finite-horizon) SOAV optimal control
under simple conditions (e.g. the plant model is controllable,
the $A$-matrix is nonsingular, and the finite alphabet for the control
includes $0$).
The obtained discrete-valued control is a piecewise constant signal,
and we prove the number of discontinuities, or {\em switching times},
is bounded.
This property is very important in particular for networked control
since the upper bound of the number of switching times,
which can be given before optimization, ensures the upper bound
of the data rate required to transmit the discrete-valued control.

We also prove that the value function, 
which is defined as the optimal value of the cost function of
the optimal control problem,
is a continuous and convex function of initial states.
This property is applied to prove the stability of the
model predictive control (MPC) feedback system
based on the finite-horizon SOAV optimal control.
As mentioned above, the SOAV optimal control
can be obtained by the fast ADMM algorithm,
and hence the control is well-adapted for
MPC.

The remainder of this paper is organized as follows: 
In Section~\ref{sec:math}, we give mathematical preliminaries for our subsequent discussion.
In section~\ref{sec:optimal}, we formulate optimal control problem so that 
optimal controls have the desired discrete values.
After that, we examine optimal controls, 
and lead the existence, discreteness, and uniqueness of the
SOAV optimal control.
A numerical optimization algorithm based on ADMM is also presented in this section.
Section~\ref{sec:value} investigates the continuity and the convexity of the value function in SOAV optimal control.
Section~\ref{sec:MPC} gives the model predictive control formulation
and shows the stability.
Section~\ref{sec:example} presents an example of
model predictive control
to illustrate the effectiveness of the proposed method.
In Section~\ref{sec:conclusion}, we offer concluding remarks.
%
%
%
%


 

\section{Mathematical preliminaries}
\label{sec:math}
This section reviews basic definitions, facts, and notation that will be used
throughout the paper.

Let $n$ be a positive integer.
For a vector $x\in{\mathbb{R}}^n$
and a scalar $\varepsilon>0$, 
the {\em $\varepsilon$-neighborhood} of $x$ is defined by
${\mathcal B}(x,\varepsilon)\triangleq\{y\in\mathbb{R}^n: \|y-x\|<\varepsilon\}$, 
where $\|\cdot\|$ denotes the Euclidean norm in ${\mathbb{R}}^n$.
Let ${\mathcal X}$ be a subset of ${\mathbb{R}}^n$.
A point $x\in {\mathcal X}$ is called an {\em interior point} of ${\mathcal X}$ if there exists $\varepsilon>0$ 
such that ${\mathcal B}(x,\varepsilon)\subset {\mathcal X}$. 
The {\em interior} of ${\mathcal X}$ is the set of all interior points of ${\mathcal X}$, 
and we denote the interior of ${\mathcal X}$ by $\mathrm{int}{\mathcal X}$.
A set ${\mathcal X}$ is said to be {\em open} if ${\mathcal X}=\mathrm{int}{\mathcal X}$. 
A point $x\in\mathbb{R}^n$ is called an {\em adherent point} of ${\mathcal X}$ 
if ${\mathcal B}(x,\varepsilon)\cap {\mathcal X} \neq\emptyset$ for every $\varepsilon>0$, 
and the {\em closure} of ${\mathcal X}$, denoted by $\overline{{\mathcal X}}$, 
is the set of all adherent points of ${\mathcal X}$.
A set ${\mathcal X}\subset{\mathbb{R}}^n$ is said to be {\em closed} if 
${\mathcal X}=\overline{{\mathcal X}}$.
The {\em boundary} of ${\mathcal X}$, denoted by $\partial {\mathcal X}$,
is the set of all points in the closure of ${\mathcal X}$, 
not belonging to the interior of ${\mathcal X}$, 
i.e., $\partial {\mathcal X}= \overline{{\mathcal X}}-\mathrm{int}{\mathcal X}$, 
where $\mathcal{X}_1-\mathcal{X}_2$ is the set of all points that belong to the set ${\mathcal X}_1$ 
but not to the set ${\mathcal X}_2$. 
In particular, if ${\mathcal X}$ is closed, then $\partial {\mathcal X}= {\mathcal X} - \mathrm{int} {\mathcal X}$, 
since ${\mathcal X}=\overline{{\mathcal{X}}}$.
A set ${\mathcal X}\subset{\mathbb R}^n$ is said to be {\em convex} if,
for any $x,y\in{\mathcal X}$ and any $\lambda\in[0,1]$,
$(1-\lambda)x+\lambda y$ belongs to ${\mathcal{X}}$.

A real-valued function $f$ defined on $\mathbb{R}^n$ is said to be 
{\em lower semi-continuous} on $\mathbb{R}^n$ if for every $\alpha\in\mathbb{R}$ the set
$\{x\in \mathbb{R}^n: f(x)>\alpha\}$
is open. 
It is known that if 
a function $f$ is lower semi-continuous on $\mathbb{R}^n$,
then 
\[
 f(x)\leq \liminf_{y \to x} f(y)
\]
for every $x\in \mathbb{R}^n$ \cite[pp. 32]{Cla}.
A real-valued function $f$ defined on a convex set ${\mathcal{C}}\subset{\mathbb{R}^n}$ 
is said to be {\em convex} if 
\[
 f\bigl((1-\lambda)x+\lambda y\bigr)\leq(1-\lambda)f(x)+\lambda f(y),
\] 
for all $x$, $y\in \mathcal{C}$ and all $\lambda\in(0,1)$.

Let $T>0$. For a continuous-time signal $u(t)$ over a time interval $[0, T]$, 
we define its {\em $L^1$ and $L^{\infty}$ norms} respectively by
\[
\|u\|_{1}\triangleq\int_{0}^{T}|u(t)| dt,
\mbox{ and } \|u\|_{\infty}\triangleq\esssup_{t\in[0, T]}|u(t)|.
\]
We denote by $m$ the Lebesgue measure on ${\mathbb{R}}$.

\section{Discrete-valued Control Problem}
\label{sec:optimal}
In this paper, we consider a linear time-invariant system represented by
\begin{equation}
\dot{x}(t)=Ax(t)+Bu(t), \quad t\geq 0
\label{eq:S}
\end{equation}
where $x(t)\in{\mathbb{R}}^n$, $u(t)\in{\mathbb{R}}$, $A\in{\mathbb{R}}^{n\times n}$,
and $B\in{\mathbb R}^{n\times 1}$.
We here assume single-input control for simplicity.
For the system \eqref{eq:S}, 
we assume {\em discrete-valued control},
that is, the control $u(t)$ can only take values in
a fixed finite set (or finite alphabet)
\begin{equation}
 {\mathbb{U}} \triangleq \{\pm U_1, \pm U_2, \dots, \pm U_{N}\}
 \label{eq:U}
\end{equation}
where $U_1,\dots,U_{N}$ are non-negative real numbers satisfying
\begin{equation}
 0\leq U_{\min}=U_1 < U_2 <  \dots < U_{N}=1.
 \label{def_Uk}
\end{equation}
Here we assume the maximum value $U_{N}=1$ without loss of generality
(otherwise, use $B/U_{N}$ instead of $B$ in \eqref{eq:S}).
Let an initial state $\xi\in{\mathbb{R}}^n$ and a finite time $T>0$ are given.
The control objective is to obtain a discrete-valued control $u(t)\in{\mathbb{U}}$
for $t\in[0,T]$ that steers the state $x(t)$ from the initial state $\xi$ 
to the origin at time $T$.
We will show in this paper that such a discrete-valued control
can be efficiently obtained by
{\em sum-of-absolute-values} (SOAV) optimal control described below.

SOAV optimal control is an extended version of
$L^1$ optimal control (also known as {\em minimum-fuel control}
\cite{AthFal}).
Let denote by ${\mathcal U}(\xi)$ the set of all {\em feasible controls}
that satisfy $x(0)=\xi$, $x(T)=0$, and $\|u\|_\infty\leq 1$
for the system \eqref{eq:S}.
We assume that ${\mathcal U}(\xi)$ is non-empty.
This assumption is satisfied if $T$ is greater than the minimum time $T^\star$
of the time optimal control \cite{HerLas}.
Then the $L^1$ optimal control is a control that minimizes the $L^1$ cost function
$\|u\|_1$ among all feasible $u\in{\mathcal{U}}(\xi)$.
It is known that the $L^1$ optimal control takes only $0$ and $\pm 1$ 
when the system \eqref{eq:S} is {\em normal},
that is, the coefficient matrix $A$ is non-singular and the pair $(A, B)$ is controllable \cite[Theorem 6-13]{AthFal}.
In other words, if \eqref{eq:S} is normal, then the $L^1$ optimal control 
gives a discrete-valued control on ${\mathbb{U}}$ with
$U_1=0$ and $U_2=1$ ($N=2$).
To extend this idea to a general set ${\mathbb{U}}$ as in \eqref{eq:U},
we consider the following SOAV cost function:
\begin{equation}
 J(u)=\sum_{i=1}^{N}w_i\phi_i(u),\quad \phi_i(u)\triangleq \|u-U_{i}\|_{1}+\|u+U_{i}\|_{1}
 \label{eq:Ju}
\end{equation}
where $w_1,\dots,w_N$ are weights satisfying $w_1+w_2+\cdots+w_{N}=1$.
The motivation for this cost function is based on the observation that
if $u(t)=U_i$ on a set ${\mathcal{I}}\subset [0,T]$,
then $u(t)-U_i=0$ on ${\mathcal{I}}$, which is {\em sparse} and
reduces the $L^1$ norm $\|u-U_i\|_1$ as discussed in \cite{NagQueNes16}.

Let us formulate the associated optimal control problem as follows.
\begin{problem}[SOAV optimal control problem]
\label{prob:ex-L1}
For a given initial state $\xi\in\mathbb{R}^n$, 
find a feasible control $u\in\mathcal{U}(\xi)$ that minimizes
the SOAV cost function
$J(u)$ given in \eqref{eq:Ju}.
\end{problem}

We will show that under some assumptions on the system \eqref{eq:S}
and the initial state $\xi$,
the SOAV optimal control takes its values in the set ${\mathbb{U}}$.

\subsection{Existence}
Here we show the existence theorem for the SOAV optimal control.

Let us define the {\em reachable set} of initial values from which
the state $x(t)$ in \eqref{eq:S} is steered to the origin
by some control $u(t)$, $t\in[0,T]$ with $\|u\|_\infty\leq 1$.
\begin{definition}[reachable set]
For the system \eqref{eq:S}, the reachable set ${\mathcal{R}}$ at time $T$ is defined by
\[
 \mathcal{R}\triangleq\bigg\{\int_{0}^{T}e^{-At}Bu(t)dt: \|u\|_{\infty}\leq 1\bigg\} \subset \mathbb{R}^n.
\]
\end{definition}
Then we have the following existence theorem.
\begin{theorem}[existence]
\label{th:exist}
For each initial state in the reachable set $\mathcal{R}$, there exists an SOAV optimal control.
\end{theorem}
\begin{IEEEproof}
Let an initial state $\xi\in \mathcal{R}$ be fixed.
The feasible control set ${\mathcal{U}}(\xi)$ can be described by
\begin{equation}
 {\mathcal{U}}(\xi) = \biggl\{u\in L^1: \int_0^T e^{-At}Bu(t)dt=-\xi, \|u\|_\infty\leq 1\biggr\}.
 \label{eq:setU}
\end{equation}
Since the set $\mathcal{U}(\xi)$ is non-empty,
we can define 
\[\theta\triangleq\inf\{J(u):u\in \mathcal{U}(\xi)\}.\]
Then there exists a sequence $\{u_l\}_{l\in{\mathbb{N}}}\subset{\mathcal{U}}(\xi)$
such that $\lim_{l\rightarrow\infty}J(u_l)=\theta$,
$\|u_l\|_\infty\leq 1$,
and
\begin{equation}
 \xi=-\int_{0}^{T}e^{-At}Bu_l(t)dt.
 \label{eq:ad}
\end{equation}
Since the set 
$\{u\in L^\infty:\|u\|_\infty\leq 1\}$
is sequentially compact in the $\mbox{weak}^{\ast}$ topology of $L^{\infty}$
\cite[Theorem A.9]{MarTeoDln},
there exist a measurable function $u_{\infty}$ 
with $\|u\|_\infty\leq 1$
and a subsequence $\{u_{l'}\}$ such that $\{u_{l'}\}$ 
converges to $u_{\infty}$ in the $\mbox{weak}^{\ast}$ topology of $L^{\infty}$,
that is, we have
\begin{equation}
 \lim_{l'\rightarrow\infty}\int_0^T \bigl(u_{l'}(t)-u_\infty(t)\bigr)f(t)dt = 0
 \label{eq:weak-star}
\end{equation} 
for any $f\in L^1$.
Since \eqref{eq:ad} satisfies for $l=l'$,
we have 
\[
 \xi=-\int_{0}^{T}e^{-At}Bu_{\infty}(t)dt
\]
and hence $u_{\infty}\in{\mathcal{U}}(\xi)$. 
Put 
\begin{equation}
 J_{l'}^{\pm} \triangleq
  \sum_{i=1}^{N}w_i  \int_{0}^{T}
    (u_{l'}(t)\pm U_i) \mbox{sgn}(u_{\infty}(t)-U_i)dt
 \label{eq:Anpm}
\end{equation}
where the function sgn is defined by
\[\mbox{sgn}(v)=
\begin{cases}
v/|v|, & \text{~if~} v\neq0,\\
0, & \text{~if~} v=0.
\end{cases}
\]
From \eqref{eq:weak-star}, we have
\[
\lim_{l'\rightarrow\infty} J_{l'}^{\pm}
 =\sum_{i=1}^{N}w_i \|u_{\infty}\pm U_i\|_1.
\]
Let $J_{l'}\triangleq J_{l'}^{+}+J_{l'}^{-}$. Then the above equation gives
\begin{equation}
\lim_{l'\rightarrow\infty} J_{l'} 
 = \lim_{l'\rightarrow\infty} (J_{l'}^{+}+J_{l'}^{-})
 =J(u_{\infty}).\label{An'to}
\end{equation}
Also, from \eqref{eq:Ju} and \eqref{eq:Anpm}, we have
\begin{equation}
J_{l'}\leq |J_{l'}|\leq\sum_{i=1}^{N}w_i \phi_i(u_{l'}) =J(u_{l'})
\label{An'leq}
\end{equation}
for each $l'\in\mathbb{N}$. 
Since the sequence $\{J(u_l)\}$ converges to $\theta$ as $l\rightarrow \infty$, 
the subsequence $\{J(u_{l'})\}$ has the same limit $\theta$.
Therefore we have 
\begin{equation}
J(u_{\infty}) = \lim_{l'\to\infty} J_{l'} \leq \lim_{l'\to\infty}J(u_{l'})=\theta
\label{ineq:existence_cost}
\end{equation}
from (\ref{An'to}) and (\ref{An'leq}).
On the other hand, since $u_{\infty}\in{\mathcal{U}}(\xi)$,
we have $J(u_{\infty})\geq\theta$.
This with \eqref{ineq:existence_cost}, we have $J(u_{\infty})=\theta$,
and $u_{\infty}$ is an optimal control for the initial state $\xi$.
\end{IEEEproof}

\subsection{Discreteness of SOAV optimal control}
Here we show the SOAV optimal solution is a discrete-valued control on ${\mathbb{U}}$.
The following theorem is one of the main results.
\begin{theorem}[discreteness]
\label{optimalcontrol}
Assume that the coefficient matrix $A$ is non-singular and the pair $(A, B)$ is controllable.
If an SOAV optimal control $u^{\ast}$ exists,
then either of the followings holds.
\begin{enumerate}
\item[(i)] $u^{\ast}(t)\in{\mathbb{U}}$ for almost all $t\in[0, T]$.
\item[(ii)] $\|u^{\ast}\|_{\infty}\leq U_{\min}$.
\end{enumerate}
In particular, if $U_{\min}=0$, then
$u^{\ast}(t)$ takes a value in ${\mathbb{U}}$
for almost all $t\in[0, T]$.
\end{theorem}
\begin{IEEEproof}
The Hamiltonian $H$ for the SOAV optimal control problem is defined by
\[
 H(x, u, p)\triangleq L(u)+p^{\mathrm{T}}(Ax+Bu)
\]
where
\[
 L(u)\triangleq\sum_{i=1}^{N}w_i(|u-U_i|+|u+U_i|)
\]
and $p$ is the costate vector.
Let $x^{\ast}$ denote the trajectory corresponding to $u^{\ast}$.
From Pontryagin's minimum principle \cite{AthFal}, 
there exists a costate vector $p^{\ast}$ satisfying 
$H(x^{\ast}, u^{\ast}, p^{\ast})\leq H(x^{\ast}, u, p^{\ast})$,
or
\begin{equation}
 L(u^{\ast})+(p^{\ast})^{\mathrm{T}}Bu^{\ast} \leq L(u)+(p^{\ast})^{\mathrm{T}}Bu
\label{hamilton-ineq}
\end{equation}
for every $u$ with $|u|\leq 1$.
Therefore, the optimal control is the minimizer of the right hand side of (\ref{hamilton-ineq}),
which can be obtained analytically as follows.

An elementary computation yields
\begin{equation}
L(u)=
\begin{cases}
-a_{k}u+b_{k},&\mbox{if }u\in[-U_{k+1}, -U_{k}],\\
2\sum_{i=1}^{N}w_iU_i,&\mbox{if }u\in[-U_{\min}, U_{\min}],\\
a_{k}u+b_{k},&\mbox{if }u\in[U_{k}, U_{k+1}]
\end{cases}
\label{L-integrand}
\end{equation}
for $k=1,2,\ldots,N-1$, where 
\begin{equation}
a_k\triangleq 2\sum_{i=1}^{k}w_i,\quad b_k\triangleq2\sum_{i=k+1}^{N}w_iU_i, \quad k=1, 2, \dots, N-1.
\label{eq:akbk}
\end{equation}
Put 
$q(t)\triangleq p^{\ast}(t)^{\mathrm{T}}B\in\mathbb{R}$,
$f(u)\triangleq L(u)+q u$, and
\[
 c_k \triangleq 
 \begin{cases}
  -a_{N-k}+q, &k=1,\dots,N-1,\\
 q, & k=N,\\
 a_{k-N}+q, &k=N+1,\dots,2N-1.
 \end{cases}
\]
From \eqref{L-integrand} and \eqref{eq:akbk},
it is easily shown that $f(u)$ is continuous and $c_1<c_2<\dots<c_{2N-1}$.
Then we have the following.
\begin{enumerate}
\item  If $c_1>0$, then
\[\argmin_{|u|\leq 1} f(u)=-U_{N}=-1.\]

\item If $c_k<0$ and $c_{k+1}>0$ for $k\in \{1, 2, \dots, 2N-2\}$, then we have 
\[
 c_1<\cdots<c_k<0<c_{k+1}<\cdots<c_{2N-1}.
\]
This implies that
\[
 \argmin_{|u|\leq 1} f(u) =
  \begin{cases}
   -U_{N-k}, & k=1,\dots,N-1,\\
   U_{\min}, & k=N,\\
   U_{k-N+1}, &k=N+1,\dots,2N-2.
  \end{cases}
\] 

\item If $c_{2N-1}<0$, then we have
\[
 c_1<c_2<\cdots<c_{2N-1}
\]
and hence
\[\argmin_{|u|\leq 1} f(u)=U_{N}=1.\]

\item If $c_k=0$ for $k\in\{1, 2,\dots, 2N-1\}$, then we have
\[
 \begin{split}
 &\argmin_{|u|\leq1} f(u)\\
 &\in
 \begin{cases}
  [-U_{N-k+1},-U_{N-k}], &k=1,\dots,N-1,\\
  [-U_{\min},U_{\min}], &k=N,\\
  [U_{k-N},U_{k-N+1}], &k=N+1,\dots,2N-1.
 \end{cases}
 \end{split}
\] 
In this case, the minimizer of $f(u)$ is not determined uniquely.
\end{enumerate}
In summary, 
the minimizer of $f(u)$, that is the SOAV optimal control $u^{\ast}$, is given by
\[
u^{\ast}(t)=
\begin{cases}
-1,&\mbox{if } a_{N-1}<q(t),\\
-U_{N-k},&\mbox{if } a_{N-k-1}<q(t)<a_{N-k},\\
-U_{\min},&\mbox{if } 0<q(t)<a_{1},\\
U_{\min},&\mbox{if } -a_1<q(t)<0,\\
U_{k+1},&\mbox{if } -a_{k+1}<q(t)<-a_{k},\\
1,&\mbox{if } q(t)<-a_{N-1}
\end{cases}
\]
where $k=1,2,\dots,N-2$ and
\begin{equation}
 u^{\ast}(t) \in
 \begin{cases}
  [-U_{N-k+1}, -U_{N-k}], &\mbox{if } q(t)=a_{N-k},\\
  [-U_{\min}, U_{\min}], &\mbox{if }q(t)=0,\\
  [U_{k}, U_{k+1}], &\mbox{if } q(t)=-a_{k}
 \end{cases}
 \label{eq:u_singular}
\end{equation} 
where $k=1,2,\dots,N-1$.

Next we claim that 
\begin{equation}
 m\bigl(\{t\in[0, T]: q(t)=\pm a_{k}\}\bigr)=0
\label{singular_pm}
\end{equation}
for every $k\in\{1, 2, \dots, N-1\}$,
where $m$ denotes the Lebesgue measure.
Let $k\in\{1, 2, \dots, N-1\}$ be fixed, and assume 
$m(\{t\in[0, T]: q(t)= a_{k}\})>0$.
Then we have
\begin{equation}
 q(t)=p^{\ast}(t)^{\mathrm{T}}B= a_{k}\label{eq:singular1}
\end{equation}
on a set $E\subset[0,T]$ with $m(E)>0$.
From Pontryagin's minimum principle, we have
\begin{equation}
{\dot p}^{\ast}(t)=-A^{\mathrm{T}}p^{\ast}(t)\label{eq:pontdif}
\end{equation}
for $t\in[0, T]$,
and hence we have 
$p^{\ast}(t)^{\mathrm{T}}AB=0$
for $t\in E$ by differentiating (\ref{eq:singular1}).
Again, by differentiating this equation,
we also have
$p^{\ast}(t)^{\mathrm{T}}A^2B=0$
for $t\in E$ from (\ref{eq:pontdif}).
Repeating this yields 
$p^{\ast}(t)^{\mathrm{T}}A^{l}B=0$
on $E$ for every $l\in\mathbb{N}$.
Therefore we have
\begin{equation}
 p^{\ast}(t)^{\mathrm{T}} A \begin{bmatrix}B & AB &\dots & A^{n-1}B\end{bmatrix}=0
 \label{eq:singularity_condition}
\end{equation}
for $t\in E$.
Since $a_k\neq0$ for every $k\in\{1, \dots, N-1\}$,
it follows from (\ref{eq:singular1}) that 
$p^{\ast}(t)$ is not identically $0$ on $[0, T]$,
and hence the determinant of 
$A [B~AB\ldots A^{n-1}B]$
is $0$.
However, this contradicts to the assumption that
the matrix $A$ is non-singular and the pair $(A, B)$ is controllable.
Therefore
$m(\{t\in[0, T]: q(t)= a_{k}\})=0$ holds
for every $k\in\{1, \dots, N-1\}$.
Similarly, we can also prove that
$m(\{t\in[0, T]: q(t)= -a_{k}\})=0$ for every $k\in\{1, 2, \dots, N-1\}$,
and hence \eqref{singular_pm} holds for every $k\in\{1, 2, \dots, N-1\}$.

Next, let assume
$m(\{t\in[0, T]: q(t)=0\})>0$.
Then we have 
$p^{\ast}(t)^{\mathrm{T}}B=0$
on a set $F\subset[0,T]$.
From (\ref{eq:pontdif}), by a similar computation as above,
we have the relation \eqref{eq:singularity_condition} for $t\in F$.
Since the matrix $A [B~AB\ldots A^{n-1}B]$ is non-singular from the assumption, 
it follows that 
\begin{equation}
p^{\ast}(t)=0,\quad \forall t\in F.
\label{peq0}
\end{equation}
Since we have 
$p^{\ast}(t)=e^{-A^{\mathrm{T}}t}p_0$
on $[0, T]$ for some $p_0\in\mathbb{R}^n$ from (\ref{eq:pontdif}), 
it follows from (\ref{peq0}) that $p_0=0$,
and hence $p^{\ast}(t)=0$ on $[0, T]$.
Then $q(t)=p^{\ast}(t)^{\mathrm{T}}B=0$ on $[0, T]$,
and we have 
$\|u^{\ast}\|_{\infty}\leq U_1=U_{\min}$
from (\ref{eq:u_singular}).
Therefore, if $q(t)$ is $0$ on a set with positive measure, 
then the optimal control $u^{\ast}$ satisfies $\|u^{\ast}\|_{\infty}\leq U_{\min}$, 
and otherwise, the optimal control takes discrete values $\pm U_1, \dots, \pm U_{N}$ 
for almost all $t\in[0, T]$.
\end{IEEEproof}

Theorem \ref{optimalcontrol} suggests that
if $U_{\min}=0$ then the SOAV optimal control is a discrete-valued control that
takes values in ${\mathbb{U}}$.
Otherwise,
it is useful to derive a condition for the optimal control $u^{\ast}$ to satisfy 
the statement (i) in Theorem \ref{optimalcontrol}.
In fact, it will be shown that there exists a subset of ${\mathbb{R}}^n$
such that if an initial state $\xi$ is in this set
then the SOAV optimal control takes values in ${\mathbb{U}}$.
To derive such a subset, we prepare the following lemmas. 
\begin{lemma}
\label{minimumofcost}
The cost function $J$ has the minimum value 
\[
 J_{\min}\triangleq 2T\sum_{i=1}^{N}w_iU_i
\]
on the set $\{u\in L^1: \|u\|_{\infty}\leq 1\}$.
\end{lemma}
\begin{IEEEproof}
See Appendix \ref{appendix1}.
\end{IEEEproof}

\begin{lemma}
\label{Rmin_reachable}
Let ${\mathcal{R}}_{\min}$ be the set of all initial values at which the optimal value is equal to $J_{\min}$.
Then we have
\[
 \mathcal{R}_{\min}=\bigg\{\int_{0}^{T}e^{-At}Bu(t)dt: \|u\|_{\infty}\leq U_{\min}\bigg\}.
\]
In particular, if $U_{\min}=0$, then we have $\mathcal{R}_{\min}=\{0\}$.
\end{lemma}
\begin{IEEEproof}
See Appendix \ref{appendix2}.
\end{IEEEproof}

Now let us state the following theorem on the discreteness of
the SOAV optimal control.
\begin{theorem}[discreteness for nonzero $U_{\min}$]
\label{thm:shape_general}
Assume that $A$ is non-singular and the pair $(A, B)$ is controllable.
If $\xi\in \mathcal{R} - \mathcal{R}_{\min}$, 
then the optimal controls take values in ${\mathbb{U}}$ almost everywhere in $[0,T]$.
Otherwise, if $\xi\in \mathcal{R}_{\min}$,
then the optimal controls take values
less than or equal to $U_{\min}$ on $[0, T]$.
\end{theorem}
\begin{IEEEproof}
This follows from Theorem \ref{th:exist}, Theorem \ref{optimalcontrol}, 
Lemma \ref{minimumofcost} and Lemma \ref{Rmin_reachable}, immediately.
\end{IEEEproof}

Next, we show the uniqueness theorem of the SOAV optimal control.
\begin{theorem}[uniqueness]
\label{th:uniqueness}
Assume that $A$ is non-singular and the pair $(A, B)$ is controllable.
Then the SOAV optimal control for the initial state $\xi\in \mathcal{R}-\mathcal{R}_{\min}$ is unique.
In particular, if $U_{\min}=0$, the SOAV optimal control is unique for any initial state $\xi\in \mathcal{R}$.
\end{theorem}
\begin{IEEEproof}
Fix an initial state $\xi\in \mathcal{R}-\mathcal{R}_{\min}$, 
and let $u_1$ and $u_2$ be optimal controls for $\xi$.
Then we have 
\begin{equation}
J(u_1)=J(u_2)\leq J(u)\label{eq:uniq1}
\end{equation}
for all $u\in\mathcal{U}(\xi).$
For any $\lambda\in(0, 1)$, 
the control $\lambda u_1+(1-\lambda)u_2$ is feasible for $\xi$,
and hence the convexity of the cost function $J$ and \eqref{eq:uniq1} yields
\[
J(u_1)\leq J(\lambda u_1+(1-\lambda)u_2)\leq \lambda J(u_1)+(1-\lambda)J(u_2)
=J(u_1)
\]
which means $ J(\lambda u_1+(1-\lambda)u_2)=J(u_1)$.
Therefore $\lambda u_1+(1-\lambda)u_2$ is an optimal control for $\xi$.

Put
\begin{align*}
  &E_1 = \{t\in[0, T]: u_1(t)\in\mathbb{U}\},\\
  &E_2 = \{t\in[0, T]: u_2(t)\in\mathbb{U}\},\\
  &F    =  \{E_1\cap E_2: u_1(t)= u_2(t)\},\\
  &G   =\{E_1\cap E_2: u_1(t)\neq u_2(t)\}.
\end{align*}
From theorem \ref{thm:shape_general}, 
we have $m(E_1)=m(E_2)=T$, and then we also have $m(E_1\cap E_2)=T$.

Here, there exist some $\lambda\in(0, 1)$ such that 
$\lambda u_1(t) +(1-\lambda)u_2(t)\not\in \mathbb{U}$ for any $t\in G$,
from the gap between the uncountability of $(0, 1)$ and the countability of $\mathbb{U}$.
It follows from 
the optimality of the control $\lambda u_1+(1-\lambda)u_2$ for such $\lambda$ and Theorem \ref{thm:shape_general} 
that $m(G)=0$.
Therefore we have
\begin{align*}
  m(F)
  = m(F) + m(G)
  = m(E_1\cap E_2)
  = T,
\end{align*}
and then
\[
  T=m(F)\leq m(\{t\in[0, T]: u_1(t)=u_2(t)\})\leq T,
\]
which yields 
\[m(\{t\in[0, T]: u_1(t)=u_2(t)\}) = T.\]
This means the uniqueness.
\end{IEEEproof}

From the above discussion, the SOAV optimal control can give
a discrete-valued control taking values in ${\mathbb{U}}$
under some assumptions on the system \eqref{eq:S} and the initial state $\xi$.
A discrete-valued control is a piecewise constant signal,
and changes its value at switching instants.
It is undesirable for real applications if the number of switching were infinite, however it never happens. In fact, 
we have the following theorem that gives an upper bound of the number of switching.
\begin{theorem}[number of switching]
Assume that $A$ is non-singular and the pair $(A, B)$ is controllable.
Then the number $M$ of switching
of the SOAV optimal control for each initial state $\xi\in \mathcal{R}-\mathcal{R}_{\min}$ satisfies
\begin{equation}
M<n(2N-1)(1+\Omega T/\pi),
\label{eq:number_of_switching}
\end{equation}
where $\Omega$ is the largest imaginary part of the eigenvalues of $A$.
In particular, if $U_{\min}=0$, 
then the number $M$ of switching for each initial state $\xi\in \mathcal{R}$ satisfies
\begin{equation}
 M<2n(N-1)(1+\Omega T/\pi).
 \label{eq:number_of_switching2}
\end{equation}
\end{theorem}
\begin{IEEEproof}
Fix arbitrarily an initial state $\xi\in \mathcal{R}-\mathcal{R}_{\min}$, 
and let $u^{\ast}$ be the SOAV optimal control for the initial state $\xi$.
From the proof of Theorem \ref{optimalcontrol}, 
$u^{\ast}(t)$ has discontinuities at points 
such that $q(t)=p^{\ast}(t)^\mathrm{T}B \in S\triangleq \{0, \pm a_{1}, \dots, \pm a_{N-1}\}$,
where $a_k$ is defined in \eqref{eq:akbk}.
Take arbitrarily an element $a\in S$.
Since $p^{\ast}(t)=e^{-A^{\mathrm{T}}t}p_{0}$ for some $p_{0}\in\mathbb{R}^n$,
we have $q(t)=p_{0}^{\mathrm{T}}e^{-At}B$.
Therefore $q(t)=a$ implies 
$a-p_{0}^{\mathrm{T}}e^{-At}B=0$, or
\[
\begin{bmatrix}a & -p_{0}^{\mathrm{T}}\end{bmatrix} 
\exp\biggl(\begin{bmatrix}0&0\\0&A\end{bmatrix}t\biggr)
\begin{bmatrix}1\\ B\end{bmatrix}=0.
\]
The number of zeros on $[0,T]$ of the function on the left hand side
is less than $n(1+T\Omega/\pi)$
according to \cite{Haj77}.
Counting the elements of the set $S=\{0, \pm a_{1}, \dots, \pm a_{N-1}\}$ yields the estimate
\eqref{eq:number_of_switching}.
In particular, if $U_{\min}=0$, 
the switching instants of the SOAV optimal control for $\xi\in \mathcal{R}$ consist of all $t$ such that 
$q(t)\in\{\pm a_{1}, \dots, \pm a_{N-1}\}$,
and the estimate \eqref{eq:number_of_switching2} holds.
\end{IEEEproof}

\section{Value function}
\label{sec:value}
In this section, we investigate the value function in the SOAV optimal control.
The value function is the optimal value of the SOAV optimal control, defined by
\[
 V(\xi)\triangleq\mathrm{min}\{J(u):u\in \mathcal{U}(\xi)\},\quad\xi\in \mathcal{R}
\]
where $J(u)$ is defined in \eqref{eq:Ju}.
From the existence theorem (Theorem \ref{th:exist}), this is well-defined.
In this section, 
we will show the continuity of the value function $V(\xi)$.
This property plays an important role to prove the stability 
when the optimal control is extended to model predictive control
(see Section \ref{sec:MPC} below).
To prove the continuity, 
the following lemmas are fundamental.
\begin{lemma}
The value function $V(\xi)$ is convex on $\mathcal{R}$.
\label{lem:convex}
\end{lemma}
\begin{IEEEproof}
See Appendix \ref{appendix:lem:convex}.
\end{IEEEproof}

\begin{lemma}
\label{closedness}
For $\alpha\geq J_{\min}$, 
let 
\[\mathcal{R}_{\alpha}\triangleq\bigg\{\int_{0}^{T}e^{-At}Bu(t)dt: 
\|u\|_{\infty}\leq 1, J(u)\leq\alpha\bigg\}.\]
Then the set $\mathcal{R}_{\alpha}$ is closed for every $\alpha\geq J_{\min}$.
\end{lemma}
\begin{IEEEproof}
See Appendix \ref{appendix3}.
\end{IEEEproof}

\begin{lemma}
\label{boundary}
If the pair $(A, B)$ is controllable, 
then we have
\[\mathcal{R}=\{\xi:V(\xi)\leq 2T\}.\]
In particular, 
\[\partial \mathcal{R}=\{\xi:V(\xi)=2T\}.\]
\end{lemma}
\begin{IEEEproof}
See Appendix \ref{appendix4}.
\end{IEEEproof}

From these lemmas, we show the continuity of the value function $V(\xi)$
based on the discussion given in \cite{IkeNag16}.
\begin{theorem}[continuity of $V$]
If the pair $(A, B)$ is controllable,
then $V(\xi)$ is continuous on $\mathcal{R}$.
\label{continuity}
\end{theorem}
\begin{IEEEproof}
Define
\[
\overline{V}(\xi)\triangleq
\begin{cases}
V(\xi),&\xi\in \mathcal{R},\\
2T,&\xi\in\mathbb{R}^n-\mathcal{R}.
\end{cases}
\]
It is sufficient to show that $\overline{V}(\xi)$ is continuous on $\mathbb{R}^n$.

First, we show that $\overline{V}(\xi)$ is continuous at every 
$\xi\in\partial\mathcal{R}$.
Fix $\xi\in\partial \mathcal{R}$.
For $\alpha$ that satisfies $\alpha\geq 2T$ or $\alpha<J_{\min}$, 
the set
$\{\xi: \overline{V}(\xi)>\alpha\}$
is empty or $\mathbb{R}^n$, respectively.
For $\alpha$ with $J_{\min}\leq\alpha<2T$,
we have 
\[
 \{\xi: \overline{V}(\xi)>\alpha\}=\mathbb{R}^n-\{\xi: \overline{V}(\xi)\leq\alpha\}
\]
which is open from Lemma \ref{closedness} since 
\[\{\xi: \overline{V}(\xi)\leq\alpha\}=\{\xi: {V}(\xi)\leq\alpha\}=\mathcal{R}_{\alpha}.\]
It follows that the set $\{\xi: \overline{V}(\xi)>\alpha\}$ is open for every real number $\alpha$, 
and hence $\overline{V}(\xi)$ is lower semi-continuous on $\mathbb{R}^n$.
Then we have
\begin{equation}
\overline{V}(\xi)\leq\liminf_{\eta\to\xi}\overline{V}(\eta).
\label{lowerconti}
\end{equation}
On the other hand, we have
\begin{equation}
\limsup_{\eta\to\xi}\overline{V}(\eta)\leq 2T,
\label{upperconti}
\end{equation}
from Lemma \ref{boundary}.
Therefore,
\[\overline{V}(\xi)\leq\liminf_{\eta\to\xi}\overline{V}(\eta)\leq\limsup_{\eta\to\xi}\overline{V}(\eta)\leq 2T=\overline{V}(\xi)\]
from (\ref{lowerconti}), (\ref{upperconti}), and Lemma \ref{boundary}.
This yields
\[
 \overline{V}(\xi)=\lim_{\eta\to\xi}\overline{V}(\eta)
\]
which means that $\overline{V}(\xi)$ is continuous at every $\xi\in\partial\mathcal{R}$.

Since $V(\xi)$ is convex on $\mathcal{R}$ from Lemma \ref{lem:convex} 
and $\mathcal{R}$ contains the origin in its interior 
from the controllability of the pair $(A, B)$ \cite[Theorem 17.3, Corollary 17.1]{HerLas},
$V(\xi)$ is continuous at every point in $\mbox{int}\mathcal{R}$ \cite[Theorem 10.1, p.44]{Roc}.

Therefore $\overline{V}(\xi)$ is continuous on $\mathbb{R}^n$, and then $V(\xi)$ is continuous on $\mathcal{R}$.
\end{IEEEproof}

\section{An extension to MPC}
\label{sec:MPC}
In this section, we extend the finite-horizon SOAV optimal control discussed above
to infinite-horizon model predictive control (MPC).

Suppose that we are given a sequence $\{t_{k}\}_{k\in{\mathbb{N}}}$ of sampling instants. 
We assume that 
\begin{equation}
 0=t_{0}<t_{1}<t_{2}<\cdots
 \label{eq:sampling_instants1}
\end{equation}
and there exists $\tau>0$ such that 
\begin{equation}
 \tau\leq t_{k+1}-t_{k}\leq T,\quad k=0,1,2,\dots
  \label{eq:sampling_instants2}
\end{equation}
We also assume that the initial state $x(0)=x_0\in{\mathcal{R}}$ is given.
At each sampling instant $t_k$, 
the SOAV optimal control, say $u_k$, with horizon length $T$ is computed
by solving the SOAV optimal control problem (Problem \ref{prob:ex-L1})
with $\xi=x(t_k)$.
We then apply $u_k(t)$ on the $k$-th time interval $[t_{k}, t_{k+1}]$.
If each optimization has the optimal solution, then this process gives a control
\begin{equation}
u(t)=u_k(t-t_k),\quad t\in[t_k, t_{k+1}],\quad k=0,1,2,\dots
\label{eq:u_mpc}
\end{equation}
From the assumptions \eqref{eq:sampling_instants1} and \eqref{eq:sampling_instants2}, the control $u(t)$ is defined for all $t\in[0, \infty)$.

\subsection{Stability}
Here we investigate the stability of the closed-loop system
with the model predictive control given in \eqref{eq:u_mpc}.
More precisely, 
the question here is whether the origin is stable in the sense of Lyapunov regardless of the choice
of sampling instants $\{t_k\}$ with the control \eqref{eq:u_mpc}.

Note that the state $x(t)$ for $t\in[0, t_1]$ obviously exists in the reachable set ${\mathcal{R}}$ 
while the control $u_0$ is used, 
since every point out of the set ${\mathcal{R}}$ needs a time duration more than 
$T$ to be steered to the origin by any control $v$ with $\|v\|_{\infty}\leq 1$.
Therefore, we have $x(t_1)\in{\mathcal{R}}$, and 
the next optimal control $u_1$ exists on the next interval $[t_1,t_2]$.
Then the state $x(t)$ for $t\in[t_1, t_2]$ lies in the reachable set ${\mathcal{R}}$ 
while the control $u_1$ is used.
It follows that the state $x(t)$ lies in the reachable set ${\mathcal{R}}$ 
for all $t\in[0,\infty)$ under this situation 
and each optimization has the optimal solution, 
and hence the control $u$ is well defined.
Then the continuity of the value function (Theorem \ref{continuity}) leads to the stability of the closed-loop system,
as described in the following theorem.
\begin{theorem}[stability]
\label{MPC}
If the pair $(A, B)$ is controllable and $U_{\min}=0$,
then the origin is stable in the sense of Lyapunov regardless of 
the choice of the sampling instants $t_0,t_1,\dots$ that satisfy
\eqref{eq:sampling_instants1} and \eqref{eq:sampling_instants2}
when we use the control $u$ defined in \eqref{eq:u_mpc}.
\end{theorem}
\begin{IEEEproof}
Fix a sequence $\{t_k\}_{k=0}^{\infty}$ of sampling instants that satisfy
\eqref{eq:sampling_instants1} and \eqref{eq:sampling_instants2}.
Also fix a positive real number $\varepsilon>0$.
We can take a real number $r\in(0, \varepsilon)$ such that 
\[
 \mathcal{B}_r\triangleq\{\xi\in \mathbb{R}^n: \|\xi\|\leq r\}\subset \mathcal{R}
\]
since $\mathcal{R}$ contains the origin in its interior from the controllability of the pair $(A, B)$.
From Theorem \ref{continuity}, $V(\xi)$ is continuous on $\partial \mathcal{B}_r$,
and then we can define 
\[\alpha\triangleq \min_{\|\xi\|=r} V(\xi).\]
From Lemma \ref{Rmin_reachable}, we have $V(\xi)>J_{\min}$ for the initial state $\xi\neq0$,
and hence $\alpha>J_{\min}$.
Take $\beta\in(J_{\min}, \alpha)$.
Then the set $\mathcal{R}_{\beta}\cap \partial \mathcal{B}_r$ is empty,
and $\mathcal{R}_{\beta}$ contains the origin and is convex.
Hence we have $\mathcal{R}_{\beta}\subset \mbox{int} \mathcal{B}_r$.
From the continuity of $V(\xi)$ at the origin, 
there exists $\delta>0$ such that $\|\xi\|\leq\delta$ implies 
\begin{equation}
J_{\min}\leq V(\xi)\leq\beta.
\label{ineq:mpc}\end{equation}
When we use the control $u$ defined in \eqref{eq:u_mpc} for $\xi$ with $\|\xi\|\leq\delta$,
it is clear that we have 
\begin{equation}
V(x_{\xi}(t))\leq V(\xi),\quad \forall t\geq 0
\label{ineq:mpc_2}
\end{equation}
where $x_{\xi}(t)$ is the state with $x_{\xi}(0)=\xi$ and is obtained by using $u$.
Therefore for $\xi$ with $\|\xi\|\leq\delta$ we have $V(x_{\xi}(t))\leq \beta$ for all $t\geq0$ from \eqref{ineq:mpc} and \eqref{ineq:mpc_2}.
Since $\mathcal{R}_{\beta}\subset \mathcal{B}_{r}\subset \mathcal{B}_{\varepsilon}$, 
for any initial state $\xi$ with $\|\xi\|\leq\delta$ 
we have $x_{\xi}(t)\in \mathcal{B}_{\varepsilon}$ for all $t\geq0$,
which means that the origin is stable in the sense of Lyapunov.
\end{IEEEproof}

\subsection{Numerical Optimization}
\label{subsec:ADMM}
Here we propose a numerical computation algorithm
to solve the (finite-horizon) SOAV optimal control problem 
to obtain a discrete-valued control input.

For simple plants, such as single or double integrators,
the discrete-valued control can be obtained in a closed form
via Pontryagin's minimum principle as the discussion in
\cite[Chap.~8]{AthFal} for $L^1$ optimal control.
However, for general linear time-invariant plants,
one should rely on numerical computation.
For this, we adopt a time discretization approach
to solve the SOAV control problem.
This approach is standard for numerical optimization;
see e.g.~\cite[Sec.~2.3]{Ste}.
We then derive
an algorithm for the optimization based on
the {\em alternating direction method of multipliers} (ADMM)
\cite{ADMM1,DRS2,ADMMBoyd}.
This algorithm is simple but much faster than the standard
interior point method.

We first divide the interval $[0,T]$ into $\nu$ subintervals,
$[0,T] = [0,h) \cup \dots \cup [(\nu-1)h,\nu h]$,
where $h$ is the discretization step chosen such that $T=\nu h$.
We here assume (or approximate) that the state $x(t)$ and the control $u(t)$ are
constant over each subinterval.
On the discretization grid,
$t=0,h,\dots,\nu h$,
the continuous-time plant \eqref{eq:S} is described as
\[
 x_d[l+1] = A_d x_d[l] + B_d u_d[l], \quad l=0,1,\dots,\nu-1
\]
where $x_d[l]\triangleq x(lh)$, $u_d[l]\triangleq u(lh)$, and
\[
 A_d \triangleq e^{Ah},\quad B_d \triangleq\ \int_0^h e^{At}Bdt.
\]
Set the control vector
\[
 z \triangleq \bigl[u_d[0], u_d[1],\dots,u_d[\nu-1]\bigr]^{\mathrm T}.
\] 
Let $\xi$ be the initial state, that is, $x(0)=\xi$.
Then the final state $x(T)$ can be described as
\[
 x(T)=x_d[\nu]=\zeta + \Phi z
\]
where $\zeta \triangleq A_d^\nu \xi$ and
\[
 \Phi \triangleq \begin{bmatrix}A_d^{\nu-1}B_d,&A_d^{\nu-2}B_d,&\dots,&B_d\end{bmatrix} \in{\mathbb{R}}^{n\times\nu}.
\]
Rename the discrete values in ${\mathbb{U}}$ as
\[
 r_1\triangleq -U_N, r_2\triangleq -U_{N+1}, \dots,
r_{L}\triangleq U_{N}
\]
where $L\triangleq 2N$ and the weights for $J(u)$ in \eqref{eq:Ju} as
\[
 p_1=p_{L}\triangleq w_N, p_2=p_{L-1}\triangleq w_{N-1}, \dots,
p_{N}=p_{N+1}\triangleq w_1.
\]
Then the SOAV optimal control problem is approximated by
\begin{equation}
 \begin{aligned}
  & \underset{z \in{\mathbb{R}}^{\nu}}{\text{minimize}}
  & & \sum_{i=1}^L p_i \|z-r_i\|_{\ell^1}\\
  & \text{subject to}
  & & \|z\|_{\ell^\infty}\leq 1,~
  \Phi z + \zeta = 0,
 \end{aligned}
 \label{eq:SOAV}
\end{equation}
where $\|\cdot\|_{\ell^1}$ and $\|\cdot\|_{\ell^\infty}$
are the $\ell^1$ and $\ell^\infty$ norms in
${\mathbb{R}}^\nu$, respectively.
The optimization problem \eqref{eq:SOAV}
is reducible to linear programming \cite{Nag15}, and can be solved by standard
numerical software packages, such as \verb=cvx= with MATLAB
\cite{cvx,gb08}, based on the interior point method.
However, for large scale problems, the computational burden of
such an algorithm becomes heavy,
and hence we give a more efficient algorithm based on ADMM.

\subsubsection{Alternating direction method of multipliers (ADMM)}
We here briefly review the ADMM algorithm.
The ADMM
solves the following type of convex optimization.
\begin{equation}
 \begin{aligned}
  & \underset{z \in{\mathbb{R}}^{N_1}, y \in{\mathbb{R}}^{N_2}}{\text{minimize}}
  & & f(z) + g(y)\\
  & \text{subject to}
  & & y=\Psi z
 \end{aligned}
 \label{eq:ADMM}
\end{equation}
where $f:\R^{N_1}\mapsto\R\cup\{\infty\}$ and $g:\R^{N_2}\mapsto\R\cup\{\infty\}$ are
proper lower semi-continuous convex functions, and $\Psi\in\R^{N_2\times N_1}$.
The algorithm of ADMM is given, for $y[0], w[0]\in\R^{N_2}$ and $\gamma>0$, by
\begin{equation}
 \left\{
 \begin{split}
z[j+1] &\leftarrow\argmin_{z\in{\mathbb{R}^{N_1}}}\bigl\{f(z)+\tfrac{1}{2\gamma} \bigl\| y[j]-\Psi z- w[j]\bigr\|^2\bigr\}\\
y[j+1] &\leftarrow\prox_{\gamma g} \bigl(\Psi z[j+1]+ w[j]\bigr)\\ 
w[j+1] &\leftarrow w[j]+\Psi z[j+1] - y[j+1]
 \end{split}
\right.
\label{alg:ADMM}
\end{equation}
for $j=0,1,2,\dots$,
where $\prox_{\gamma g}$ denotes the {\em proximity operator} of $\gamma g$ defined by
\begin{equation*}
\prox_{\gamma g}(z)\triangleq \argmin_{y\in\R^{N_2}}\gamma g(y)+\frac{1}{2}\|z-y\|^2.
\end{equation*}


We recall a convergence analysis of ADMM by Eckstein-Bertsekas \cite{DRS2}.
\begin{theorem}[Convergence of ADMM \cite{DRS2}]\label{fact:ADMMconv}
Consider the optimization problem~\eqref{eq:ADMM}.
Assume that $\Psi^{\mathrm{T}} \Psi$ is invertible
and that a saddle point of its unaugmented Lagrangian
$\Lmath_0(z,y,w)\triangleq f(z)+g(y)- (\Psi z-y)^{\mathrm{T}}w$
exists.
Then the sequence $\{(z[j],y[j])\}_{j\in\mathbb{N}}$ generated by Algorithm~\eqref{alg:ADMM} 
converges to a solution of \eqref{eq:ADMM}.
\end{theorem}

\subsubsection{Reformulation into ADMM-applicable form}
In what follows, we reformulate our optimization problem
described in \eqref{eq:SOAV} into the standard form in \eqref{eq:ADMM} to apply ADMM.

Let $\Omega_1 \triangleq \{z\in\R^\nu|\;\| z \|_{\ell^\infty}\leq1\}$ be the unit-ball of the infinity norm,
and $\Omega_2\triangleq \{-\zeta\}$ be the singleton consisting of the vector $-\zeta$.
Define the indicator function of a nonempty closed convex set by
\begin{equation*}
\iota_{\Omega}(z)\triangleq
\begin{cases}
0,&\mbox{if }z\in \Omega,\\
\infty,&\mbox{otherwise.}
\end{cases}
\end{equation*}
Then, we can rewrite the optimization problem \eqref{eq:SOAV} as
\begin{equation}\label{eq:SOAV2}
\underset{z \in{\mathbb{R}}^{\nu}}{\text{minimize}} \sum_{i=1}^Lp_i\|z-r_i\|_{\ell^1}
+\iota_{\Omega_1}(z)+\iota_{\Omega_2}(\Phi z).
\end{equation}
Introducing new variables $y_1,\dots,y_{L+2}$ such that
$y_i=z$ ($i=1,\dots,L+1$), and $y_{L+2}=\Phi z$,
we can translate \eqref{eq:SOAV2} into
\begin{equation}
 \begin{aligned}
 & \underset{z\in{\mathbb{R}}^{N_1},y\in{\mathbb{R}}^{N_2}}{\text{minimize}}
 & & \sum_{i=1}^Lp_i\|y_i-r_i\|_{\ell^1}+\iota_{\Omega_1}(y_{L+1})+\iota_{\Omega_2}(y_{L+2})\\
 & \text{subject to}
 & & y=\Psi z
 \end{aligned}
 \label{eq:SOAV3}
\end{equation}
where $N_1\triangleq \nu$, $N_2\triangleq (L+1)\nu+n$,
$y\triangleq [y_1^{\mathrm{T}} \dots y_{L+2}^{\mathrm{T}}]^{\mathrm{T}}\in\R^{N_2}$, and
\begin{equation*}
\Psi\triangleq 
 \begin{bmatrix}I &\dots &I & \Phi^{\mathrm{T}}\end{bmatrix}^{\mathrm{T}}\in\R^{N_2\times N_1}.
\end{equation*}

Finally, by setting
\begin{align*}
f(z)&\triangleq0,\\
g(y)&\triangleq \sum_{i=1}^Lp_i\|y_i-r_i\|_{\ell^1}+\iota_{\Omega_1}(y_{L+1})+\iota_{\Omega_2}(y_{L+2})
\end{align*}
the optimization problem \eqref{eq:SOAV3} is reduced to the standard from of \eqref{eq:ADMM}.

\subsubsection{Computation}
Since $f=0$, the first step of \eqref{alg:ADMM} becomes strictly convex quadratic minimization,
which boils down to solving linear equations, that is,
\begin{align*}
z[j+1]&=\argmin_{z\in{\mathbb{R}}^\nu}\tfrac{1}{2\gamma}\|y[j]-\Psi z-w[j]\|^2\\
&=(\Psi^{\mathrm{T}} \Psi)^{-1} \Psi^{\mathrm{T}}(y[j]-w[j])\\
&=\bigl((L+1)I+\Phi^{\mathrm{T}}\Phi\bigr)^{-1}v[j]
\end{align*}
where
\begin{equation*}
v[j]\triangleq\sum_{i=1}^{L+1}(y_i[j]-w_i[j])+\Phi^{\mathrm{T}}(y_{L+2}[j]-w_{L+2}[j]).
\end{equation*}
Note that the inverse matrix $\bigl((L+1)I+\Phi^{\mathrm{T}}\Phi\bigr)^{-1}$ can be computed
off-line.

On the other hand, the second step of \eqref{alg:ADMM} can be separated with respect to each $y_i$.
For $y_i$ ($i=1,\ldots,L$), we have to compute the proximity operator of the $\ell_{1}$ norm with shift $r_i$,
which is reduced to a simple soft-thresholding operation:
for $l=1,\ldots,\nu$,
\begin{align*}
&\bigl[\prox_{\gamma p_i\|\cdot-r_i\|_{1}}(z)\bigr]_{(l)}
 = r_i + \prox_{\gamma p_i|\cdot|}({z_{(l)}}-r_i)\\
 &=r_i+\mbox{sgn}(z_{(l)}-r_{i(l)})\max\{|z_{(l)}-r_{i}|-\gamma p_i,0\}
\end{align*}
where $(\cdot)_{(l)}$ denotes the $l$-th entry of a vector.
Here we use the shift property of the proximity operator (see, e.g., \cite{TechCombettes}).

For $y_{L+1}$ and $y_{L+2}$, the computation of the proximity operators of the indicator functions are required.
Since the proximity operator of the indicator function of a nonempty closed convex set $\Omega$ 
equals to the metric projection $P_{\Omega}$ onto $\Omega$,
the updates of $y_{L+1}$ and $y_{L+2}$ are reduced to calculating $P_{\Omega_1}$ and $P_{\Omega_2}$, respectively.
We can compute $P_{\Omega_1}$ as follows:
\begin{equation*}
P_{\Omega_1}(z)\triangleq
\begin{cases}
z, & \mbox{if }\|z\|_{\ell^\infty}\leq 1,\\
\tilde z,& \mbox{otherwise}
\end{cases}
\end{equation*}
where
\begin{equation*}
\tilde z \triangleq \bigl[\mbox{sgn}(z_{(1)})\min\{|z_{(1)}|,1\}\dots \mbox{sgn}(z_{(\nu)})\min\{|z_{(\nu)}|,1\}\bigr]^{\mathrm{T}}.
\end{equation*}
Meanwhile, $P_{\Omega_2}=P_{\{-\zeta\}}$ is simply give by
\begin{equation*}
P_{\Omega_2}(z) \triangleq -\zeta.
\end{equation*}

As addressed in \cite{ADMMBoyd}, ADMM tends to converge to modest accuracy within a few tens of iterations.
This property is favorable in real-time control systems.

\section{Example}
\label{sec:example}
\begin{figure}[t]
  \centering
   \includegraphics[width=\linewidth]{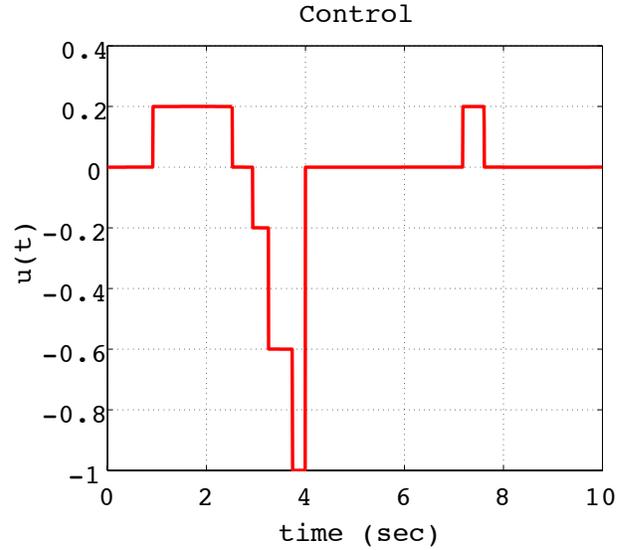}
  \caption{Discrete-valued control by SOAV MPC.}
  \label{fig:control}
\end{figure}

\begin{figure}[t]  
  \centering
   \includegraphics[width=\linewidth]{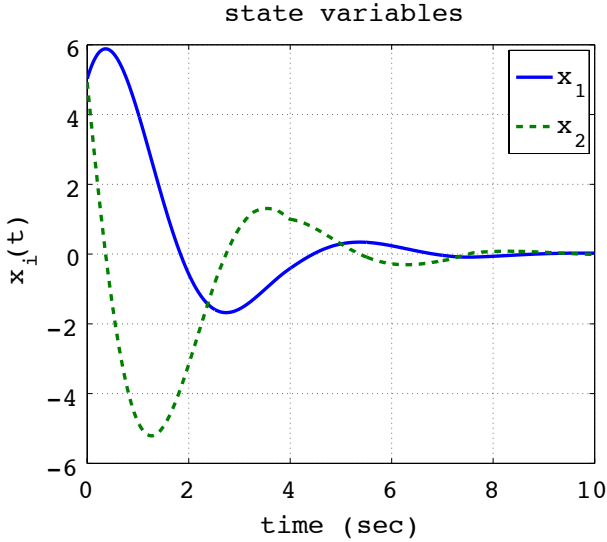}
  \caption{State $x(t)=[x_1(t),x_2(t)]^{\mathrm{T}}$}
  \label{fig:state}
\end{figure}

In this section, 
we give an example of model predictive control
based on the SOAV optimal control.
Let us consider the plant model represented in
\[
\dot{x}(t)
=
\begin{bmatrix}
0&1\\
-2&-1
\end{bmatrix}
x(t) +
\begin{bmatrix}
0\\
1
\end{bmatrix}
u(t), \quad t\geq 0.
\]
For this system, we consider the following SOAV optimal control problem:
\begin{align*}
&\mbox{minimize} \quad J(u)=\sum_{i=1}^{4} w_i \phi_i(u)\\
&\mbox{subject to} \quad x(0)=[5, 5]^{\mathrm{T}},\quad x(5)=0,\quad \|u\|_{\infty}\leq 1
\end{align*}
where $w_i=0.1i$ ($i=1,2,3,4$),
and $U_1=U_{\min}=0$, $U_2=0.2$, $U_3=0.6$ and $ U_4=1$.
The sampling instants are taken as $t_1=4$, $t_2=8$, $t_3=9$ and $t_4=10$.
Fig.~\ref{fig:control} shows the control $u$ defined by \eqref{eq:u_mpc} 
and Fig.~\ref{fig:state} shows the state trajectory according to $u$.
Certainly, 
we can see that the control $u$ takes only discrete values $0$, $\pm 0.2$,
$-0.6$ and $-1$,
and the state converges to the origin.

\section{Conclusion} 
\label{sec:conclusion}
In this paper,
we have proposed sum-of-absolute-values (SOAV) optimization
for discrete-valued control.
We have shown the existence and uniqueness theorems
of the SOAV optimal control.
We have also given conditions
for the SOAV optimal control to generate a
discrete-valued control signal.
The obtained discrete-valued control has a finite number of switching,
of which an upper bound has been derived.
Furthermore we have investigated the continuity of the value function,
by which the stability has been proved 
when the (finite-horizon) SOAV optimal control is extended to model predictive control (MPC).
For MPC, a fast algorithm based on ADMM is proposed.
A simulation result has been illustrated to show the effectiveness
of the proposed method.

\appendices
\section{Proof of Lemma \ref{minimumofcost}}
\label{appendix1}
First, we show that the value of the cost function $J(u)$ for each $u \in \{u\in L^1: \|u\|_{\infty}\leq 1\}$ 
is greater than or equal to $J_{\min}$, 
and then we show the minimum $J_{\min}$ is achieved by $u=0$.

Fix a control $u$ with $\|u\|_{\infty}\leq U_{N}=1$, and define
\begin{equation}
\begin{split}
&E\triangleq \{t\in[0, T]: -U_1\leq u(t)\leq U_1\},\\
&E_{k}^{+}\triangleq \{t\in[0, T]: U_k< u(t)\leq U_{k+1}\},\\
&E_{k}^{-}\triangleq \{t\in[0, T]: -U_{k+1}\leq u(t)< -U_{k}\}
\label{def_set_E}
\end{split}
\end{equation}
where $k=1, 2, \dots, N-1$.
Let $\gamma\triangleq m(E)$ and $\gamma_k^{\pm}\triangleq m(E_k^{\pm})$.
Since these sets are pairwise disjoint and satisfy 
\[
 E\cup\bigcup_{k=1}^{N-1}(E_{k}^{+}\cup E_{k}^{-})=[0, T]
\]
we have 
\begin{equation}
\gamma+\sum_{k=1}^{N-1}(\gamma_{k}^{+}+\gamma_{k}^{-})=T
\label{eq:sumofsets}
\end{equation}
from the countable additivity of the Lebesgue measure.
Let
\[
\lambda_k^{\pm} \triangleq  \pm \int_{E_k^{\pm}}u(t)dt.
\] 
An elementary computation yields 
\[
 \begin{split}
  \phi_i(u) &= \int_0^T \bigl(|u(t)-U_i|+|u(t)+U_i|\bigr)dt\\
   &= 2U_i\biggl(\gamma + \sum_{k=1}^{i-1}(\gamma_k^{+}+\gamma_k^{-})\biggr)
     + 2\sum_{k=i}^{N-1}(\lambda_k^{+}+\lambda_k^{-})
 \end{split}
\]
for $i=1,2,\dots,N$, where we define $\sum_{k=1}^0=0$ and $\sum_{k=N}^{N-1}=0$.      
Then for $k=i,i+1,\dots,N-1$, we have
\[
 \lambda_k^{\pm} = \pm \int_{E_k^{\pm}} u(t)dt \geq U_k\gamma_k^{\pm}\geq U_i\gamma_k^{\pm}.
\]
It follows from \eqref{eq:sumofsets} that
\[
 \phi_i(u) \geq 2U_i\biggl(\gamma+\sum_{k=1}^{N-1}(\gamma_k^{+}+\gamma_k^{-})\biggr)=2U_iT
\]
and hence
\[
 J(u)\geq 2T\sum_{i=1}^{N} w_iU_i=J_{\min}.
\] 
Therefore the cost function $J(u)$ takes values greater than or equal to $J_{\min}$,
and $J(u)$ attains the minimum $J_{\min}$ when $u=0$.

\section{Proof of Lemma \ref{Rmin_reachable}} 
\label{appendix2}
Fix an initial state
\[
 \xi\in \bigg\{\int_{0}^{T}e^{-At}Bu(t)dt: \|u\|_{\infty}\leq U_{\min}\bigg\}.
\]
Then there exists a control $u$ satisfying
\[
 \xi=\int_{0}^{T}e^{-At}Bu(t)dt,\quad \|u\|_{\infty}\leq U_{\min}.
\]
Since $-u$ is feasible for $\xi$ and
$-u(t)\in [-U_{\min},U_{\min}]$ for almost all $t\in [0,T]$,
we have
\[
 J(-u)=2T\sum_{i=1}^{N}w_iU_i=J_{\min}.
\]
It follows that $-u$ is an optimal control, and hence $\xi\in \mathcal{R}_{\min}$.

Conversely, take an initial state $\xi\in \mathcal{R}_{\min}$ and let $u^\ast$ denote the optimal control 
that satisfies $J(u^\ast)=J_{\min}$.
Define sets $E_{k}^{+}$, $E_{k}^{-}$ and $E$ as in the proof of Lemma \ref{minimumofcost}.
Then we can easily show that
 \[
  \int_{E_{k}^{\pm}}(\pm u^\ast(t)-U_k)dt=0
 \]
for every $k=1,$ $2,$ $\dots,$ $N-1$.
Since $u^\ast(t)-U_k$ and $-u^\ast(t)-U_k$ are positive on $E_{k}^{+}$ and $E_{k}^{-}$ for every $k$ respectively,
we have $m(E_{k}^{\pm})=0$
for every $k$.
Therefore $m(E)=T$ from \eqref{eq:sumofsets}, that is, $\|-u^\ast\|_{\infty}\leq U_1=U_{\min}$.
Also, since the control $u^\ast$ steers the initial state $\xi$ to the origin at time $T$,
we have 
\[
 \xi = \int_{0}^{T}e^{-At}B\bigl(-u^\ast(t)\bigr)dt
\]
and it follows that
\[
 \xi\in\bigg\{\int_{0}^{T}e^{-At}Bu(t)dt: \|u\|_{\infty}\leq U_{\min}\bigg\}.
\]

\section{Proof of Lemma \ref{lem:convex}}
\label{appendix:lem:convex}
Fix initial states $\xi$, $\eta\in \mathcal{R}$ and a scalar $\lambda\in(0, 1)$.
From Theorem \ref{th:exist}, 
there exist optimal controls $u_{\xi}$ and  $u_{\eta}$ for the initial states $\xi$ and $\eta$, respectively.
Then we have $\lambda\xi+(1-\lambda)\eta\in \mathcal{R}$ since $\mathcal{R}$ is convex,
and the control $\lambda u_{\xi}+(1-\lambda) u_{\eta}$ is feasible
for the initial state $\lambda\xi+(1-\lambda)\eta$.
From the convexity of $\phi_i$ in $J(u)$ (see \eqref{eq:Ju}), we have
\begin{align*}
V\bigl(\lambda\xi+(1-\lambda)\eta\bigr)
&\leq J\bigl(\lambda u_{\xi}+(1-\lambda) u_{\eta}\bigr)\\
&=\sum_{i=1}^{N}w_i\phi_i\bigl(\lambda u_{\xi}+(1-\lambda) u_{\eta}\bigr)\\
&\leq \sum_{i=1}^{N}w_i\bigl(\lambda \phi_i(u_{\xi})+(1-\lambda) \phi_i(u_{\eta})\bigr)\\
&=\lambda J(u_{\xi}) + (1-\lambda) J(u_{\eta})\\
&=\lambda V(\xi) + (1-\lambda) V(\eta).
\end{align*}

\section{Proof of Lemma \ref{closedness}}
\label{appendix3}
First, we note that the set $\mathcal{R}_{\alpha}$ is well defined for $\alpha\geq J_{\min}$ 
since $J_{\min}$ is the minimum of the cost function from Lemma~\ref{minimumofcost}. 

Fix $\alpha\geq J_{\min}$, 
and take a sequence $\{\xi_{l}\}$ in $\mathcal{R}_{\alpha}$ that converges to $\xi_{\infty}
\in {\mathbb{R}}^n$.
It is sufficient to show that $\xi_{\infty}\in \mathcal{R}_{\alpha}$.

For each $\xi_l\in {\mathcal{R}}_{\alpha}$,
there exists a control $u_l$ such that
\begin{align*}
\xi_l=\int_{0}^{T}e^{-At}Bu_{l}(t)dt,
\quad \|u_l\|_{\infty}\leq 1,
\quad J(u_l)\leq\alpha.
\end{align*}

Since the set $\{u\in L^\infty: \|u\|_\infty \leq 1\}$
is sequentially compact in the $\mbox{weak}^{\ast}$ topology of $L^{\infty}$,
there exist a measurable function $u_{\infty}$ with 
$\|u_{\infty}\|_{\infty}\leq 1$,
and a subsequence $\{u_{l'}\}$ such that 
$\{u_{l'}\}$ converges to $u_{\infty}$ in the $\mbox{weak}^{\ast}$ topology of $L^{\infty}$.
Clearly, we have
\[
 \xi_{\infty}=\int_{0}^{T}e^{-At}Bu_{\infty}(t)dt.
\]

Define $J_{l'}^{\pm}$ as \eqref{eq:Anpm} and $J_{l'}\triangleq J_{l'}^+ + J_{l'}^-$.
Then we have 
\[
 J(u_{\infty}) = \lim_{l'\to\infty} J_{l'} \leq \lim_{l'\to\infty}J(u_{l'})\leq\alpha
\]
which is verified from \eqref{An'to} and \eqref{An'leq}.
It follows that $\xi_{\infty}\in \mathcal{R}_{\alpha}$.

\section{Proof of Lemma \ref{boundary}}
\label{appendix4}
First, we show
\begin{equation}
\partial \mathcal{R}=\{\xi:V(\xi)=2T\}.
\label{eq:boundary}
\end{equation}

Fix $\xi\in\partial \mathcal{R}$, 
then the feasible control for the initial state $\xi$ is only the time optimal control, 
which is determined uniquely and takes only $\pm 1$ for almost all $t\in[0, T]$
since the pair $(A, B)$ is controllable \cite{Haj72}, \cite[Theorem 12.1]{HerLas}.
Let us denote the time optimal control by $u^\star$, 
and let $F^+$, $F^-\subset[0, T]$ be the set on which $u^\star$ takes $1$ and $-1$, respectively,
that is,
\[u^\star(t)=
\begin{cases}
1,& \mbox{if}\quad t\in F^+,\\
-1,& \mbox{if}\quad t\in F^-
\end{cases}
\]
and $m(F^+) + m(F^-)=T$.
Then we have 
\[
V(\xi)=J(u^\star)
=2\sum_{i=1}^{N}w_i\bigl(m(F^+) + m(F^-) \bigr)
=2T.
\]

Conversely,
fix an initial state $\xi\in \mathcal{R}$ such that $V(\xi)=2T$.
If $\xi\in \mbox{int}\mathcal{R}$, 
then there exist a scalar $\lambda\in[0, 1)$ and 
a vector $\eta\in\partial \mathcal{R}$ such that $\xi=\lambda\eta$.
As we proved above, we have $V(\eta)=2T$.
It follows from the convexity of $V$ that 
\[
 \begin{split}
 V(\xi)
 =V(\lambda\eta)
 &\leq \lambda V(\eta) + (1-\lambda)V(0)\\
 &=2\lambda T + (1-\lambda)V(0)
 \end{split}
\]
which yields
\begin{equation}
2T\leq V(0)\label{contra1}
\end{equation}
since $V(\xi)=2T$.

However, 
since $u=0$ is feasible for the initial state $0$,
we have 
\[2T\sum_{i=1}^{N}w_iU_i\leq V(0)\leq J(0)=2T\sum_{i=1}^{N}w_iU_i\]
from Lemma \ref{minimumofcost}.
This implies 
\begin{equation}
V(0)=2T\sum_{i=1}^{N}w_iU_i<2T\sum_{i=1}^{N}w_i=2T.
\label{contra2}\end{equation}

Thus a contradiction occurs between \eqref{contra1} and \eqref{contra2},
and hence $\xi\notin\mbox{int}\mathcal{R}$.
Since $\mathcal{R}$ is closed \cite{Haj71}, 
we have $\xi\in\partial \mathcal{R}$.

Next, we show 
\[\mathcal{R}=\{\xi:V(\xi)\leq 2T\}.\]
From \eqref{eq:boundary}, it is sufficient to show 
\begin{equation}
\mbox{int}\mathcal{R}=\{\xi:V(\xi)<2T\}.
\label{interior}
\end{equation}

First, 
fix an initial state $\xi\in\mbox{int}\mathcal{R}$,
then there exist a scalar $\lambda\in[0, 1)$ and a vector 
$\eta\in\partial \mathcal{R}$ such that $\xi=\lambda\eta$,
and $V(\eta)=2T$ from \eqref{eq:boundary}.
It follows from \eqref{contra2} that 
\[
V(\xi)
\leq \lambda V(\eta) + (1-\lambda)V(0)
=2\lambda T + (1-\lambda)V(0)
<2T.
\]

Conversely, for any initial state $\xi$ such that $V(\xi)<2T$, 
we have $\xi\in \mbox{int}\mathcal{R}$ from \eqref{eq:boundary}.
Thus \eqref{interior} follows, and the proof is completed.

\section*{Acknowledgment}
This research was supported in part by JSPS KAKENHI Grant Numbers 
26120521, 15K14006, 15H02668,and 15H06197. 

\bibliographystyle{IEEEtran}
\bibliography{IEEEabrv,main}

\end{document}